\documentclass[%
 reprint,
  superscriptaddress,
 amsmath,amssymb,amsfonts,
 aps,
 prl,
 floatfix,
 longbibliography
]{revtex4-2}

\usepackage{comment}
\usepackage[utf8]{inputenc}
\usepackage[pdftex]{graphicx} \graphicspath{{}}
\usepackage{float} \usepackage{color}
\usepackage{hyperref}
\usepackage{subfigure}
\hypersetup{
  colorlinks=true,
  linkcolor=blue,
  citecolor = blue,
  urlcolor=blue,
}

\usepackage{mathtools}

\DeclarePairedDelimiterX\braket[2]{\langle}{\rangle}{#1 \delimsize\vert #2}
\DeclarePairedDelimiterX\expval[3]{\langle}{\rangle}{#1 \delimsize\vert #2  \delimsize\vert #3}
\DeclarePairedDelimiterX\proj[2]{\delimsize\vert#1\rangle}{\langle#2\delimsize\vert}{ }
\usepackage{siunitx}

\usepackage{tikz}
\usetikzlibrary{arrows, arrows.meta}
\usepackage[T1]{fontenc}

\newcommand{\vb}[1]{\mathbf{#1}}

\begin{document}

\title{Chaotic spin dynamics of elongated spinor condensates}

\author{J. Reyes-Calderón}
\affiliation{Institut f\"ur Theoretische Physik, Leibniz Universit\"at Hannover, Appelstr. 2, 30167, Hannover, Germany} 

\author{A. Gallemí}
\affiliation{Institut f\"ur Theoretische Physik, Leibniz Universit\"at Hannover, Appelstr. 2, 30167, Hannover, Germany} 
\affiliation{Departament de Física, Universitat de les Illes Balears, C/ Valldemossa km 7.5, 07122 Mallorca, Spain}
\affiliation{Institute of Applied Computing and Community Code (IAC-3), C/ Valldemossa km 7.5, 07122 Mallorca, Spain}

\author{C. Klempt}
\affiliation{Institut f\"ur Satellitengeod\"aasie und Inertialsensorik (DLR-SI), Deutsches Zentrum f\"ur Luft- und
Raumfahrt e.V. (DLR), Callinstraße 30b, 30167 Hannover, Germany}

\author{L. Santos}
\affiliation{Institut f\"ur Theoretische Physik, Leibniz Universit\"at Hannover, Appelstr. 2, 30167, Hannover, Germany}

\begin{abstract}
Elongated spin-$1$ condensates present a highly non-trivial local magnetization dynamics, due to the  interplay between nonlinear and quantum effects stemming from the inhomogeneous density profile. This interplay results in different dynamical regimes after an initial global quench. In particular, we show that the system may display the coexistence of  markedly different dynamical domains separated by a robust interface that acts as an spatial excited-state quantum phase transition. Furthermore, the local spinor dynamics may enter a chaotic regime characterized by irregular evolution and exponential sensitivity to initial conditions. We map the universal phase diagram distinguishing regular and chaotic regimes, which may be probed in on-going experiments.  
\end{abstract}

\maketitle

%%%%%%%%%%%%%%%%%%%%%%%%%%%%%%%%%%%%%%%%%%%%

% Many-body dynamics / Spinor BECs

Ultra-cold atomic gases constitute one of the foremost experimental platforms for investigating quantum many-body dynamics under highly controllable conditions~\cite{Bloch2008}.  In particular, 
atomic Bose–Einstein condensates (BECs) have emerged as paradigmatic systems for the study of 
nonlinear far-from-equilibrium dynamics~\cite{Polkovnikov2011}, due to the inherent nonlinearity introduced 
by the inter-particle interactions. 
Spinor BECs, formed by more than two internal~(spin) states~\cite{Kawaguchi2012}, 
present especially intricate dynamics due to the interplay of superfluidity, spin degrees of freedom, and nonlinear spin-dependent interactions. In contrast to scalar BECs or binary Bose mixtures, spinor gases support coherent spin-changing collisions, which enable a broad range of dynamical phenomena, including spin-mixing oscillations~\cite{Chang2004, Schmaljohann2004}, topological defect generation~\cite{Sadler2006}, thermalization~\cite{Evrard2021}, coarsening~\cite{Williamson2016, Huh2024}, and spin turbulence and chaotic dynamics~\cite{Kronjaeger2008, Tsubota2014, Kim2024}.

% ESQPTs

Recent experiments have shown that spinor BECs are also a suitable  platform for the controlled study of excited-state quantum phase transitions~(ESQPTs)~\cite{Meyer-Hoppe2023}. Contrary to ground-state transitions, 
ESQPTs constitute a singularity in the energy spectrum that separates eigenstates with markedly different properties~\cite{Stransky2014}. 
These transitions are relevant in a wide variety of many-body quantum systems, ranging from the Lipkin-Meshkov-Glick model~\cite{Leyvraz2005}, 
and the Dicke and Jaynes-Cummings models~\cite{Perez-Fernandez2011, Brandes2013}, to microwave Dirac billiards~\cite{Dietz2013},  
and molecular bending transitions~\cite{Larese2013}. 

% Trapped condensate mixtures

Many experiments on condensate mixtures, including those on ESQPTs~\cite{Meyer-Hoppe2023}, can be well described in single-mode approximation~(SMA)~\cite{Kawaguchi2012}, in which all components share the same spatial profile. This simplifying assumption typically demands a spatial extension smaller than the spin healing length, although even in 
that case resonance effects may violate the SMA~\cite{Scherer2010,Jie2023}. 
Beyond-SMA scenarios, however, are particularly interesting, as trapped mixtures provide a versatile platform for simulating the dynamics of inhomogeneous magnetic materials. In these materials, the local magnetization dynamics, typically described by the Landau-Lifshitz equation~\cite{Landau1935,Baryakhtar2015}, results from the interplay between external magnetic field, magnetic anisotropy, and inhomogeneous magnetization. In condensate mixtures~(in contrast to typical magnetic materials) the anisotropy may be larger than the external field, and the spatial density profile of the trapped BEC results in magnetic inhomogeneity. As recently shown for far-from-equilibrium coherently coupled binary BECs~\cite{Farolfi2021}, inhomogeneity may result in the formation of unstable magnetic interfaces. 

% In this Letter

In this Letter, we discuss the rich local magnetization dynamics of quenched elongated spin-$1$ BECs, characterized by decoupled spin and density degrees of freedom, and transversal, but not axial, SMA. Our results show that 
the interplay between coherent spin-changing collisions, inhomogeneous density, and quadratic Zeeman energy, allows for exploring in on-going experiments a universal dynamical phase diagram, all the way from the axial SMA regime characterized by homogeneous excited state phases~\cite{Meyer-Hoppe2023}, to the quasi-local-density approximation regime 
in which, similar to Ref.~\cite{Farolfi2021}, unstable magnetic interfaces are formed.
Most interesting is, however, the intermediate regime, characterized by the 
coexistence of markedly different dynamical domains, separated by a robust magnetic interface that acts as a spatial ESQPT, and by the appearance of a chaotic regime, with irregular magnetization dynamics, and exponential sensitivity to initial conditions.

%%%%%%%%%%%%%%%%%%%%%%%%%%%%%%%%%%%%%%%%%%%%%%%%%%%%
% MODEL

\paragraph{Model.--} We consider a harmonically trapped spin-1 BEC, described by the energy functional~\cite{Kawaguchi2012}:
\begin{eqnarray}
E|{\psi_m}] &=& 
\int d\vb{r}\left [
\sum_{m=\pm1,0} \psi_m^*
\left( \hat H_0 + qm^2 \right) \psi_m  
+ \frac{g}{2} n^2\right ]   
\nonumber \\
&+& c\int d\vb{r}\left (\frac{1}{2}(n_{1} -n_{-1})^2 + n_0 (n_{1} + n_{-1})\right ) \nonumber \\
&+& c\int d\vb{r}\left ( 
\psi_1^*\psi_{-1}^*\psi_0^2+\mathrm{H.c.}
\right )
\end{eqnarray}
where $\psi_m(\vb{r})$ is the 
wavefunction of the $m=\pm1, 0$ component, $n_m=|\psi_m|^2$, $n=\sum_m n_m$, 
$\hat H_0=-\frac{\hbar^2}{2M} \nabla^2 + V(x)+V_\perp(y,z)$, with $M$ the mass of the atoms, $V(x)=M\omega^2x^2/2$, 
$V_\perp(y,z)=M\omega_\perp^2(y^2+z^2)/2$, and 
$q$ is the quadratic Zeeman energy. 
We consider an elongated harmonic trap, with $\omega_\perp\gg \omega$. 
The spin-independent interactions, $gn^2/2$, depend on the coupling constant $g=4\pi\hbar^2\bar{a}/M$, where 
$\bar{a}=(a_0+2a_2)/3$, with $a_S$ the scattering length for the channel of spin projection $S$. The spin-dependent interactions are characterized by the 
coupling constant $c=4\pi\hbar^2\delta a/M$, with
$\delta a=(a_2-a_0)/3$. We focus in this paper on
ferromagnetic BECs~($c<0$).

We assume $g\gg |c|$, as it is the case of $F=1$ $^{87}$Rb. As a result, $n(\vb{r})$ remains invariant during the spin dynamics, being determined by the Gross-Pitaevskii equation~(GPE): $\mu_{3D} \sqrt{n(\vb{r})} = [\hat H_{0} + g n(\vb{r}) ] \sqrt{n(\vb{r})}$, with $\mu_{3D}$ the chemical potential. 
Furthermore, we are interested in elongated BECs in which the transversal~(but not the axial) extension is smaller than the spin healing length. In this transversal-SMA regime $\psi_m(\vb{r}, t)=\sqrt{n(\vb{r})}\, \chi_m(x,t)e^{-i\mu t/\hbar}$.
For the sake of simplicity, we focus on the quasi-1D regime, $\mu\equiv\mu_{3D}-\hbar\omega_\perp \ll \hbar\omega_\perp$, in which 
$n(\vb{r})=\frac{1}{\pi l_\perp^2}e^{-(y^2+z^2)/l_\perp^2}\, \bar{n}(x)$, $l_\perp^2=\hbar/M\omega_\perp$, and 
\begin{equation}
\mu \sqrt{\bar{n}(x)} = [\hat H_{x} + g_{\mathrm{1D}} \bar{n}(x) ] \sqrt{\bar{n}(x)}, 
\label{eq:n}
\end{equation}
with 
$\hat H_{x}=-\frac{\hbar^2}{2M}\frac{\partial^2}{\partial x^2} + V(x)$ and $g_{\mathrm{1D}}\equiv \frac{g}{2\pi l_\perp^2}$. However, the qualitative features discussed below only demand the (much less restrictive) transversal-SMA regime. Similar physics is expected even if the BEC is transversally in the Thomas-Fermi~(TF) regime. 
Finally, we assume the axial TF regime~($\mu/\hbar\omega\gg 1$), such that $\bar{n}(x)=\bar{n}(0)(1-x^2/R_{\mathrm{TF}}^2)$, with $R_{\mathrm{TF}}=\sqrt{2\mu/M\omega^2}$ the TF radius. 

The spinor dynamics is determined by the coupled GPEs: $i\hbar\partial_t\psi_m = \delta E/\delta\psi_m^*$. 
Since $\partial_t (\psi_1-\psi_{-1})=0$, setting $\psi_{1}({\mathbf r},t=0)=\psi_{-1}({\mathbf r},t=0)$, reduces the set to 
two coupled equations, for $\psi_0$ and $\psi_s \equiv (\psi_{1} + \psi_{-1})/\sqrt{2}$. Integrating over the $yz$ plane, and introducing $c_{\mathrm{1D}}\equiv \frac{c}{2\pi l_\perp^2}$, the GPEs
acquire the form:
\begin{eqnarray}
i\hbar\dot\psi_0\!&=&\!\left [\hat H_{x}+g_{\mathrm{1D}}\bar{n}\right ] \psi_0 \!+\! c_{\mathrm{1D}}\!\left (|\psi_s|^2\psi_0 + \psi_s^2\psi_0^*\right )\!, 
\label{eq:eq_psi_0}
\\
i\hbar\dot\psi_s\!&=&\!\left [\hat H_{x} + q +g_{\mathrm{1D}}\bar{n}\right ] \psi_s \!+\! c_{\mathrm{1D}} \!\left (|\psi_0|^2\psi_s + \psi_0^2\psi_s^*\right )\! . 
\label{eq:eq_psi_s}
\end{eqnarray}
The results discussed below are obtained from numerical simulations of Eqs.~\eqref{eq:n}, \eqref{eq:eq_psi_0}, and \eqref{eq:eq_psi_s}.

%%%%%%%%%%%%%%%%%%%

% FIGURE 1

\begin{figure*}
    \includegraphics[width=1.00\linewidth]{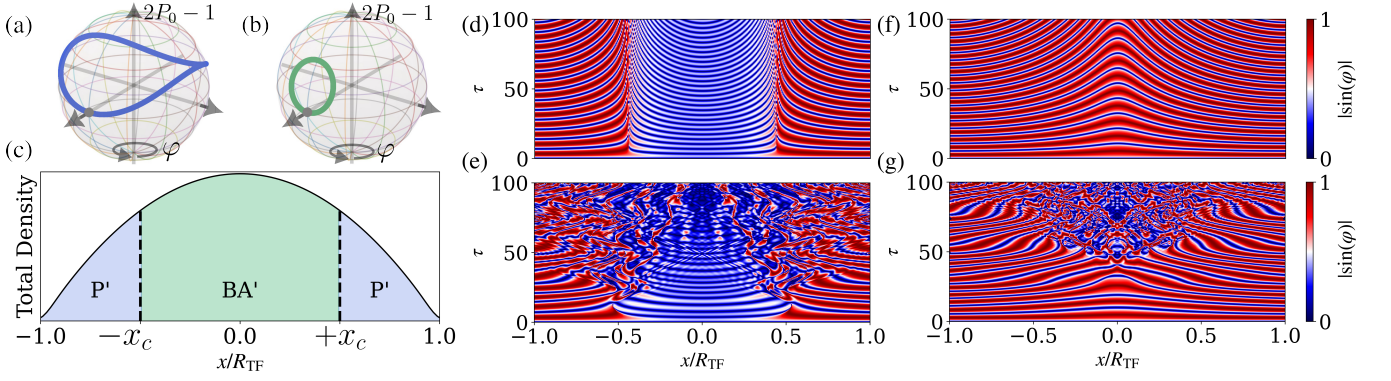}
    \caption{Typical Bloch-sphere trajectories in the P$'$ phase~(a) and the BA$'$ phase~(b). 
    (c) Sketch of of the Thomas-Fermi profile with spatial dynamical BA$'$ and P$'$ domains and an ESQPT at $x=\pm x_{c}$. Figures (d) and (e) show the spin dynamics in the LDR for $\eta=103.9$ and $\xi_0=0.45$ in the absence and the presence of quantum torque, respectively. Note the destabilization of the domain wall. Figures (f) and (g) show the same for 
    $\eta = 103.9$ and $\xi_0=0.58$. There are no domains in the LDR, but the different time scale of the local dynamics results eventually in a large quantum torque that destabilizes the LDR evolution.}
    \label{fig:figure_1}
\end{figure*}

%%%%%%%%%%%%%%%%%%%%%%%%%%%%%%%%%%%%

% SINGLE-MODE APPROXIMATION

\paragraph{Single-mode approximation.--}
If $\eta\equiv R_{TF}/l_s<1$, with $l_s\equiv \hbar/(2M|c_{\mathrm{1D}}|\bar{n}(0))^{1/2}$ the spin healing length associated to the central density, the system is also axially in the SMA regime~\cite{Feldmann2021}, $\psi_{\alpha=0,s}(x,t) = \sqrt{\bar{n}(x)}\chi_\alpha(t)e^{-i\mu t/\hbar}$, with $\chi_\alpha(t)=\sqrt{P_\alpha(t)} e^{i\phi_\alpha(t)}$, and $P_s=1-P_0$.  
For $\xi_{\mathrm{av}} \equiv \frac{q}{2|c_{\mathrm{1D}}|n_{\mathrm{av}}}<-1$,
with $n_{\mathrm{av}}=\frac{1}{N}\int dx \bar{n}(x)^2 = 4 \bar{n}(0)/5$, 
the ground state is given by $P_0=0$~(twin-Fock~(TwF) phase),  
whereas for $\xi_{\mathrm{av}}>1$, $P_0=1$~(polar~(P) phase). For 
$|\xi_{\mathrm{av}}|<1$, the ground state is in the 
broken-axisymmetric~(BA) phase, with 
$\varphi = \phi_s - \phi_0=0$, $P_0=\frac{1}{2}\left (1+\xi_{\mathrm{av}}\right )$, and energy 
$E_{GS}=-\frac{1}{2}|c_{\mathrm{1D}}| n_{\mathrm{av}} (1+\xi_{\mathrm{av}})^2$. Interestingly, for $|\xi_{\mathrm{av}}|<1$ there is an ESQPT, which can be traversed either by changing $q$ or the excitation energy $\Delta E = E-E_{GS}$, which is fixed by the initial $P_0(t=0)$ and $\varphi(t=0)$. At the ESQPT the nature of the excited states changes. For $\Delta E>\Delta E_{cr}(\xi_{\mathrm{av}})$ the  excited states are either in the TwF$'$~(for $\xi_{\mathrm{av}}<0$) or P$'$~(for $\xi_{\mathrm{av}}>0$), where $\varphi$ is a running value, and 
the Bloch vector
$\vb{v}_B(\vb{r},t)= (\theta,\varphi)$, with $2P_0-1=\cos\theta$, moves around the $z$ axis~(Fig.~\hyperref[fig:figure_1]{\ref*{fig:figure_1}(a)}). In contrast, for $\Delta E<\Delta E_{cr}(\xi_{\mathrm{av}})$ the excited states are in the BA$'$ phase, where $\vb{v}_B(\vb{r},t)$ remains localized in $\varphi$~(Fig.~\hyperref[fig:figure_1]{\ref*{fig:figure_1}(b)}), without moving around the $z$ axis.
Given $\varphi(0)=0$, $q>0$, and $|\xi_{\mathrm{av}}|<1$, the BA$'$ excited-state phase occurs for $P_0(t=0)>|\xi_{\mathrm{av}}|$, and P$'$ for $P_0(t=0)<|\xi_{\mathrm{av}}|$. Recent experiments using $F=1$ $^{87}$Rb atoms have mapped this SMA excited state phase diagram~\cite{Meyer-Hoppe2023}.

%%%%%%%%%%%%%%%%%%%%

% FIGURE 2

\begin{figure*}
    \includegraphics[width=1.00\linewidth]{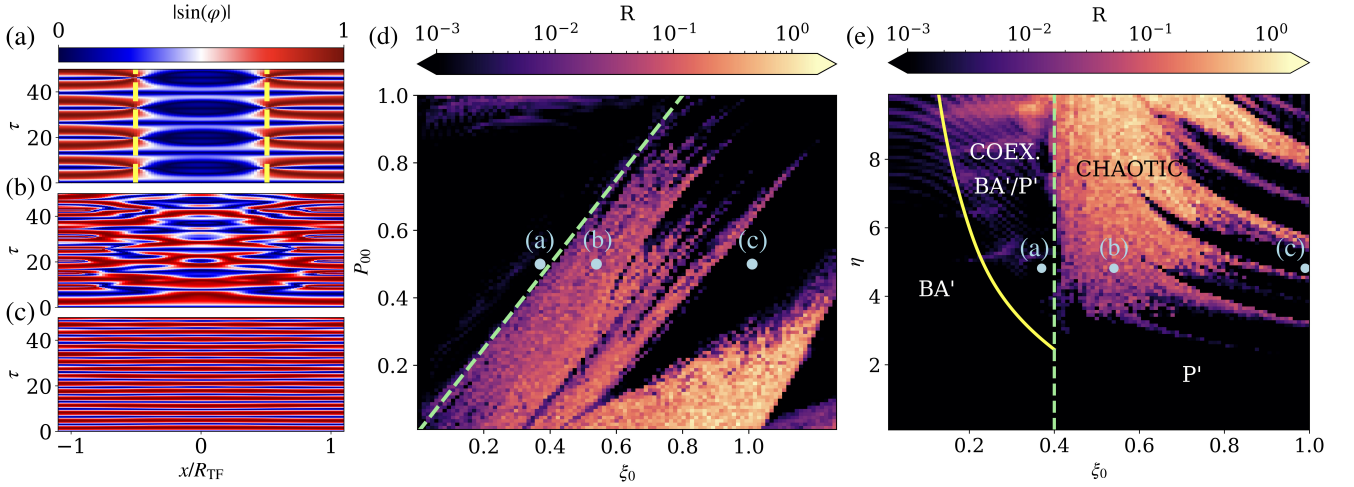}
    \caption{Dynamics of $|\sin{(\varphi(x,\tau))}|$ 
    for $P_{00}=0.5$, $\eta=4.82$, and (a) $\xi_0=0.38$~(a),  $\xi_0=0.54$~(b), and $\xi_0=1.01$~(c). 
    In Fig. (a) the yellow dashed lines indicate the ESQPT at $x=\pm x_c$.
    (d) Dynamical phase diagram as a function of $\xi_0$ and $P_{00}$, for $\eta=4.82$. (e)  Dynamical phase diagram as a function of $\xi_0$ and $\eta$ for $P_{00}=0.5$.
    In plot (d) and (e), the  
    color indicates the value of $R$~(see text).     Note the intricate structure of the chaotic regime. 
    In plot (e), the solid yellow line marks the transition between the purely BA$'$ regime, and the coexistence regime. 
    In both diagrams the dashed green lines indicate the transition between the regular coexistence regime and the irregular regime. We have also indicated in both diagrams the values employed in Figs. (a), (b) and (c).}
    \label{fig:figure_2} 
\end{figure*}

%%%%%%%%%%%%%%%%%%%%%%%%%%%%%%%%%%%%%%

% BEYOND SMA

\paragraph{Spinor dynamics beyond SMA.--} Beyond SMA~($\eta>1$), we can 
introduce the local spin $\vb{S}(x,t)=(S_x,S_y,S_z)$, 
with $S_x=\chi_0^*\chi_s+\mathrm{c.c.}$, 
$S_y = -i(\chi_0^*\chi_s-\mathrm{c.c.})$, and $S_z=|\chi_0|^2-|\chi_1|^2$. The coupled GPEs transform
into a nonlinear equation for the local spins, which closely resembles the celebrated Landau–Lifshitz equation for the local magnetization dynamics in a solid~\cite{Landau1935,Baryakhtar2015}:
\begin{eqnarray}
\partial_\tau\vb{S}&=&
-\frac{1}{2\eta^2F(x)}\partial_x \left [ 
F(x)\partial_x\vb{S}\times \vb{S}
\right ] + \vb{\Omega}\times\vb{S},
\label{eq:eq_spin}
\end{eqnarray}
with $x\equiv x/R_{TF}$, $\tau \equiv 2|c_{\mathrm{1D}}|\bar{n}(0) t$, 
$F(x) = 1-x^2$, 
$\vb{\Omega}(x,\tau)=\left ( -F(x) S_x(x,\tau), 0, -\xi_0\right )$ the local nonlinear effective magnetic field, with
$\xi_0 \equiv q/2|c_{\mathrm{1D}}|n(0)$. 
In the following, we assume that the condensate is initially prepared in an out-of-equilibrium initial spatially uniform state, characterized by $P_0(x,\tau=0)=P_{00}$ and $\varphi(x,\tau=0)=0$. The subsequent dynamics is then a universal function of  $\eta$, $\xi_0$ and $P_{00}$, being characterized by the competition between the classical torque $\vb{\Omega}(x,\tau)\times\vb{S}(x,\tau)$, and the quantum torque~(first term at the r.h.s of Eq.~\eqref{eq:eq_spin}).

%%%%%%%%%%%%%%%%%%%%%%%%%%%%%%%%%%%%%

% LDR

\paragraph{Local density regime.--} If $\eta\to\infty$, the quantum torque vanishes, and the local spinor dynamics only depends on $\bar{n}(x)$~(local density regime, LDR). The characteristic time scale is $\tau_0(x) = \hbar/\left(2|c|\bar{n}(x) \right )$. For  $\xi_0<P_{00}$, the local dynamics is BA$'$ if $\xi(x)\equiv\xi_0/F(x)<P_{00}$, i.e. if $|x|<x_c=\sqrt{1-\xi_0/P_{00}}$, and P$'$ if $\xi(x)>P_{00}$~($|x|>x_c$). Hence, in the LDR, the inhomogeneous density profile induces two distinct excited-state domains, separated by an spatial ESQPT at $x=\pm x_c$ , see Figs.~\hyperref[fig:figure_1]{\ref*{fig:figure_1}(c)} and \hyperref[fig:figure_1]{(d)}.
However, the quantum torque cannot be fully neglected even for $\eta\gg 1$. It becomes especially relevant at the domain wall at $x=x_c$, resulting in the eventual instability of the spatial ESQPT, see Fig.~\hyperref[fig:figure_1]{\ref*{fig:figure_1}(e)}. This resembles recent experiments on coherently-coupled binary Bose mixtures~\cite{Farolfi2021}, where the spin dynamics is given by a similar equation as Eq.~\eqref{eq:eq_spin}, although with a crucially different $\vb{\Omega}$, two domains form with markedly different~(self-trapped or oscillating) spin dynamics, and domain wall instability is also observed. Interestingly, the LDR evolution eventually breaks down even for $\xi_0>P_{00}$, where no domains form and the LDR dynamics is purely $P'$, due to the large quantum torque eventually resulting from the locally variable time scale $\tau_0(x)$, 
see Figs.~\hyperref[fig:figure_1]{\ref*{fig:figure_1}(f)} and \hyperref[fig:figure_1]{(g)}.

\paragraph{Coexistence regime.--} 
The opposite regime to the LDR, that with small $\eta>1$, is characterized by both the violation of the SMA and the relevant role played by the quantum torque in the whole cloud. 
For $\xi_0<\xi_{0;\mathrm{SMA}}=\frac{4}{5}P_{00}$, for which the SMA dynamics is in the BA$'$ regime, the spinor evolution, although not in SMA, remains BA$'$-like in the whole cloud for $\xi_0<\xi_{0;\mathrm{cr1}}(\eta)$~(solid yellow line in Fig.~\hyperref[fig:figure_2]{\ref*{fig:figure_2}(e)}). 
Unexpectedly, for $\xi_0>\xi_{0;\mathrm{cr1}}(\eta)$, the system, although far away from the LDR, 
presents two markedly different BA$'$ and P$'$ dynamical domains, separated by an ESQPT at $x=x_c$~(we determine $\xi_{0;\mathrm{cr1}}(\eta)$ as the value at which the dynamics of the local spin at $x=0.95$ transitions from BA$'$ to P$'$; taking other positions at the wings of the TF profile give similar results). Remarkably, in contrast to the LDR, the quantum torque favors a quasi-SMA evolution in each domain. As a result, the spatial ESQPT remains stable~(see Fig.~\hyperref[fig:figure_2]{\ref*{fig:figure_2}(a)}, where we depict the evolution of $|\sin\varphi(x,\tau)|$). For growing $\xi_0$ the central BA$'$ domain shrinks down to 
$|x|<x_{c;\mathrm{cr}}=1/\sqrt{5}$ at $\xi_0 = \xi_{0;\mathrm{SMA}}$, dashed line in Figs.~\hyperref[fig:figure_2]{\ref*{fig:figure_2}(d)} and \hyperref[fig:figure_2]{(e)}. For $\xi_0 > \xi_{0;\mathrm{SMA}}$, the SMA evolution is P$'$-like, and coexistence, which in the LDR could be expected up to $\xi_0=P_{00}$, is however not observed for growing $\eta$.

%%%%%%%%%%%%%%%%%%%%%%%%%%%%%%%%

% Chaotic regime

\paragraph{Chaotic regime.--} 
When $\xi_0>\xi_{0;\mathrm{SMA}}$, the quantum torque results for a sufficiently large $\eta$ in a very different dynamical regime, characterized by the loss of the spatial domains and a markedly irregular evolution, as illustrated in Fig.~\hyperref[fig:figure_2]{\ref*{fig:figure_2}(b)}. 
This irregular dynamics is characteristic of intermediate $\xi_0$ values, whereas for large-enough $\xi_0$ the dynamics 
is again regular, P$'$-like, see Fig.~\hyperref[fig:figure_2]{\ref*{fig:figure_2}(c)}. 
In order to characterize the irregular regime, we evaluate the evolution up to times $\tau_M=200$, and calculate the power spectrum at $x=0$:  
${\cal P}(\omega)\equiv\int_0^{\tau_M} d\tau |\sin\varphi(x=0,\tau)|e^{i\omega\tau}$. The spectrum presents discrete peaks in the regular regime, and a blurred structure in the irregular one. We characterize this by monitoring $R\equiv\left ( \int d\omega {\cal P}(\omega)^4\right )^{-1}$, which is orders of magnitude larger in the irregular regime than in the regular one. We note that other observables, including spatial averages which do not require in-situ imaging, provide similar results.

\begin{figure}
 %   \centering
    \includegraphics[width=1.0\linewidth]{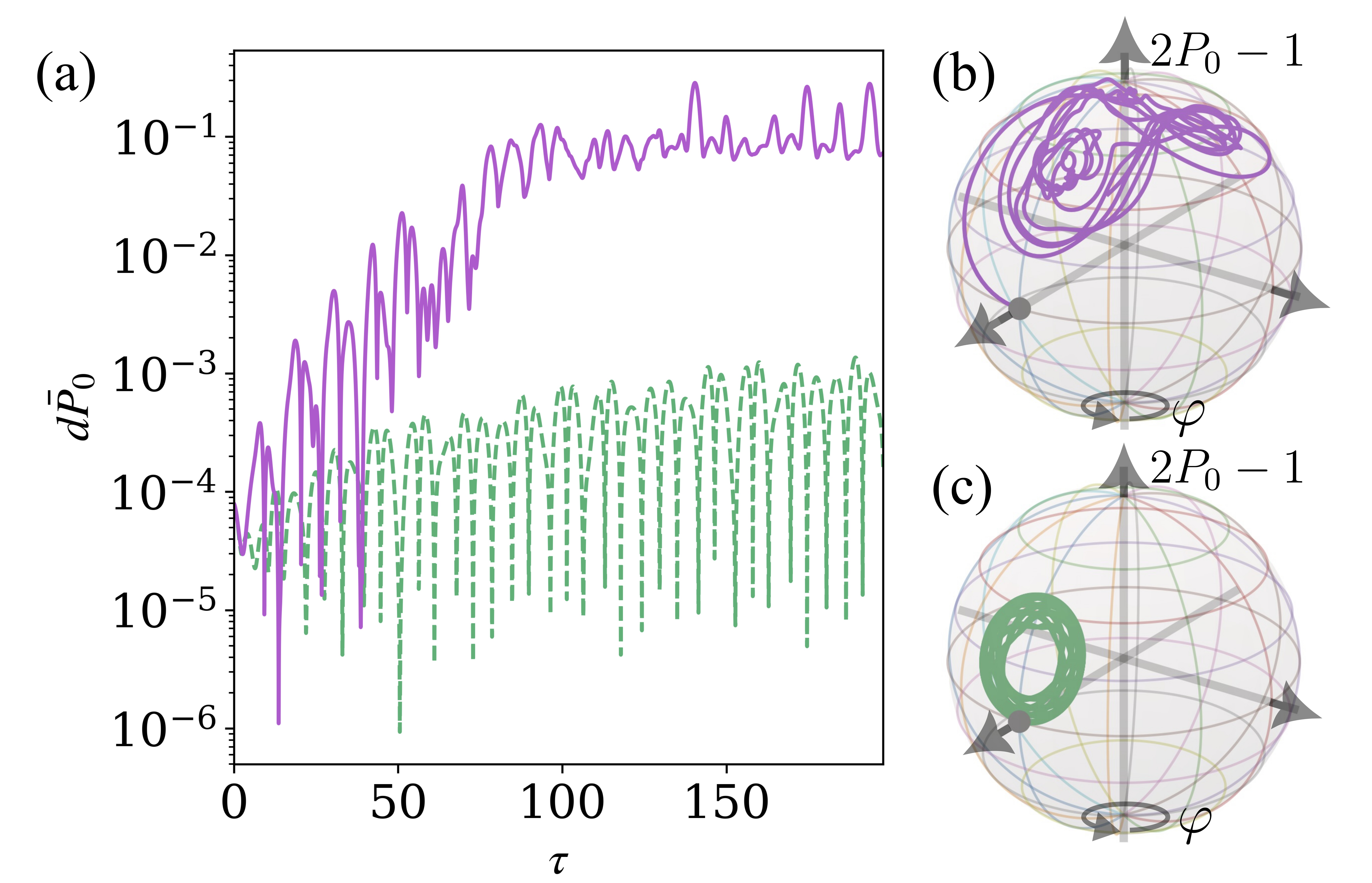}
    \caption{(a) For  $\eta = 4.82$ and $P_{00} = 0.5$, we depict the average distance $d\bar{P}_0(\tau)$~(see text) between the spatially-averaged population $\bar{P}_0(\tau)$ for an unperturbed case and a perturbed one~(initially $d\bar{P}_0(\tau=0)=10^{-4}$). For $\xi_0 = 0.49$ (irregular regime), the purple continuous line shows an exponential growth of the distance with time up to its maximum. In contrast, for $\xi_0 = 0.25$ (regular coexistence regime), the distance~(dashed green line) remains very small. The corresponding Bloch-Sphere representations for the spatially-averaged spin are depicted in (b) and (c).}
    \label{fig:figure_3}
\end{figure}

We may map in this way the chaotic and regular regimes, as illustrated in Fig.~\hyperref[fig:figure_2]{\ref*{fig:figure_2}(d)}, where we depict $R$ as a function of $\xi_0$ and $P_{00}$ for $\eta=4.82$~(corresponding e.g. to a $F=1$ $^{87}$Rb condensate with $N=2000$ atoms, $\omega_\perp/2\pi=950$~Hz, and  $\omega/2\pi=20$~Hz). In Fig.~\hyperref[fig:figure_2]{\ref*{fig:figure_2}(e)} we depict $R$ for a fixed $P_{00}=0.5$ as a function of $\eta$ and $\xi_0$. This figure nicely illustrates the different dynamical regimes when leaving the SMA regime~($\eta\lesssim 1$).  
As mentioned above, the line $\xi_0=4P_{00}/5$~(dashed line) marks the transition from the coexistence regular regime to the irregular regime. The irregular regime presents an intricate, fractal-like structure, as a function of the control parameters $\eta$, $\xi_0$ and $P_{00}$, with islands of regular dynamics within the irregular regime, as is typical in chaotic systems. We have checked the regular character of these islands~(not shown). 
The chaotic nature of the spin dynamics in the irregular regime is confirmed by monitoring the sensitivity of the system to small perturbations in initial conditions, $(P_{00},\varphi(0)) \rightarrow (P_{00},\varphi(0)) + (\delta P_{00},\delta \varphi(0))$. We compute in particular the average distance $d\bar{P}_0(\tau)$ between the spatially-averaged population $\bar{P}_0(\tau) = \int dx P_0(x,\tau) F(x)$ evaluated for the unperturbed case with that of the perturbed case. In Fig.~\hyperref[fig:figure_3]{\ref*{fig:figure_3}(a)}, we show the average distance for $\eta=4.82$, $P_{00}=0.5$. For $\xi_0=0.25$, within the regular coexistence regime, the distance remains very small in time, and for $\xi_0=0.49$, within the irregular regime, the distance grows exponentially with time up to its maximum, indicating chaotic dynamics. The corresponding evolution of the averaged spin is depicted, respectively, in Figs.~\hyperref[fig:figure_3]{\ref*{fig:figure_3}(b)} and \hyperref[fig:figure_3]{(c)}.

%%%%%%%%%%%%%%%%%%%%%%%%%%%%%%%

% FIGURE 3

%%%%%%%%%%%%%%%%%%%%%%%%%%%%%%%%%%%%%%%%%%%%%%%

\paragraph{Conclusions.--} Elongated spin-$1$ condensates present a highly non-trivial local spinor dynamics when leaving the single-mode approximation regime, characterized by the interplay between the quantum and classical non-linear torques. The system may present 
the coexistence of two different dynamical domains, separated by a domain wall that acts as an spatial excited-state quantum phase transition. 
Remarkably this occurs for a spatial condensate extension comparable to the spin healing length~(low $\eta>1$ regime), far from the local-density regime. Moreover, whereas in the local-density regime the quantum torque destabilizes the domain wall, in the 
low-$\eta$ regime it induces robust domains. We have also identified the presence of a chaotic regime, with an intricate fractal-like structure in the phase diagram, characterized by irregular spinor dynamics and exponential sensitivity to initial conditions. Although we have considered for simplicity the quasi-1D regime, these dynamical features only demand the condensate to be in the transversal-single-mode-approximation regime, and in particular the density profile could be transversally in the Thomas-Fermi regime. Finally, 
we would like to mention that, although 
fully unveiling the rich spinor dynamics may demand in-situ spin resolution, the 
regular-to-chaotic transition may be revealed as well with global spin measurements, which only demand time-of-flight experiments combined with Stern-Gerlach spin separation.

%%%%%%%%%%%%%%%%%%%%%%%%%%%%%%%%%%%%

\acknowledgments
We acknowledge financial support
from the Deutsche Forschungsgemeinschaft (DFG,
German Research Foundation), Project-ID 274200144,
SFB 1227 DQ-mat within project B01, as well as Germany’s
Excellence Strategy, EXC-2123 QuantumFrontiers,
Project-ID 390837967. 

\bibliography{references}

\end{document}